\documentclass[prd,amsfonts,nofootinbib,preprint]{revtex4}
\usepackage{amsmath,amssymb,amsfonts}
\usepackage{amsmath}
\usepackage{graphicx}
\usepackage{comment}

\usepackage{hyperref}
\hypersetup{colorlinks,linkcolor={blue},citecolor={blue},urlcolor={blue}} 
\usepackage{braket}
\usepackage[english]{babel}
\usepackage{bm}
\usepackage[utf8]{inputenc}

\usepackage{listings}

\usepackage[T1]{fontenc}
\usepackage{lmodern}
\usepackage{caption}
\usepackage{floatrow}
\usepackage{subcaption}

\usepackage{mathrsfs}
\usepackage{enumerate}

\usepackage[normalem]{ulem}
\usepackage{color}
\usepackage{array}
\def\beq{\begin{equation}}
\def\eeq{\end{equation}}
\def\be{\begin{equation}}
\def\ee{\end{equation}}
\def\bea{\begin{eqnarray}}
\def\eea{\end{eqnarray}}

\begin{document}

\title{\hfill\mbox{\small}\\[-1mm]
\hfill~\\[0mm]
       \textbf{Future leptonic $CP$ phase determination in the presence of NSI}    
       }    
%\date{}

\author{Luis A. Delgadillo}\email{ldelgadillof2100@alumno.ipn.mx
}\affiliation{Departamento de F\'{\i}sica, Escuela Superior de
F\'{\i}sica y Matem\'aticas del Instituto Polit\'ecnico Nacional\\
Unidad Adolfo L\'opez Mateos, Edificio 9, 07738 Ciudad de M\'exico, Mexico }
\author{O. G. Miranda}\email{omar.miranda@cinvestav.mx}\affiliation{Departamento de F\'{\i}sica, Centro de
  Investigaci{\'o}n y de Estudios Avanzados del IPN\\
  \textit{\small Apdo. Postal
  14-740 07000 Ciudad de M\'exico, Mexico}\\
 }
 
%% use \vspace{} (latex command) instead of \vskip (tex command) 
\vspace{0.5cm}

\begin{abstract}
\noindent
The precise determination of the leptonic $CP$-phase is one of the major goals for future generation long Baseline experiments. On the other hand, if new physics beyond the Standard Model exists, a robust determination of such a $CP$-phase may be a challenge. Moreover, it has been pointed out that, in this scenario, an apparent discrepancy in the $CP$-phase measurement at different experiments may arise. In this work, we investigate the determination of the Dirac $CP$-phase and the atmospheric mixing angle $\theta_{23}$ at several long-baseline configurations: ESSnuSB, T2HKK, and a DUNE-like experiment. We use the nonstandard neutrino interactions (NSI) formalism as a framework. We found that complementary between ESSnuSB and a DUNE-like experiment will be favorable to obtain a reliable value of the $CP$-phase, within the aforementioned scenario. Moreover, the T2HKK proposal can help to constrain the matter NSI parameters.  
\end{abstract}

\maketitle

\section{Introduction}
\label{sec:intro}

The success in neutrino physics in the last decades gives us a clear picture of the standard three oscillation parameters. The determination of the mixing angles and squared mass differences at present is remarkable, with only a few challenges to face, such as the octant problem. In this standard scenario, determining the $CP$-violating phase is the next precision physics goal and will be tackled by the next generation of long-baseline experiments (LBL). Current measurements of this important observable have been reported, mainly by NOvA and T2K collaborations. Their results seem to differ and might be a puzzle to solve if this difference persists. 
Recently, it has been pointed
out~\cite{Chatterjee:2020kkm,Denton:2020uda} that the discrepancy in
the measurement of the leptonic $CP$-violating phase
$\delta_{CP}$ experienced by the NOvA~\cite{Nova2020} and
T2K~\cite{T2K2020} experiments, can be alleviated by considering new
physics beyond the Standard Model. In particular, it has been
considered that this new physics can be parametrized in the framework
of nonstandard neutrino interactions
(NSI)~\cite{Ohlsson:2012kf,Miranda:2015dra,Farzan:2017xzy,
  Proceedings:2019qno}. Future generation of LBL experiments will
provide new $CP$ measurements with improved sensitivities that can probe
this scenario. This is the case of future proposals such as a
DUNE-like experiment~\cite{DUNE:2020jqi}, the Hyper-Kamiokande
proposal~\cite{Hyper-Kamiokande:2018ofw},
T2HKK~\cite{Hyper-Kamiokande:2016srs}, and the more recent
ESSnuSB~\cite{Alekou:2022emd}.

In general, DUNE will have a good discrimination among the
corresponding vector, scalar NSI and sterile neutrino
scenarios~\cite{Denton:2022pxt}. Furthermore, in
Ref.~\cite{Chatterjee:2021wac}, improvements on the energy resolution
at DUNE to enhance the matter NSI sensitivity were studied. It was
shown in \cite{deGouvea:2015ndi, Coloma:2015kiu}, that for sizable
NSI, DUNE will be capable of determine the off-diagonal NSI parameters
including the extra $CP$-violating phases $\phi$.

On the other hand, source and detector charged current NSI at the
European Spallation Source Neutrino Super Beam (ESSnuSB) and DUNE were
investigated in Refs.~\cite{Blennow:2015nxa, Blennow:2016etl}, it was
found that the incorporation of a near detector improves the
sensitivity to NSI at ESSnuSB, therefore competitive limits are
achieved. Besides, the authors of Ref.~\cite{Capozzi:2023ltl} explored
the benefits of a near detector at ESSnuSB, which refines the
sensitivity to matter NSI. For instance, the authors of
Ref.~\cite{Liao:2016orc} studied the implications of matter NSI
(sensitivities, degeneracies, determination) at separate neutrino
long-baseline (LBL) experiments including DUNE and the
Tokai-to-Hyper-Kamiokande-and-Korea (T2HKK) proposal.

In this work, we explore NSI effects on matter, within the framework examined in~\cite{Chatterjee:2020kkm, Denton:2020uda}, considering the effect of one flavor changing matter NSI parameter at a time, as well as the inclusion of all the matter NSI parameters, we focus on both the electron neutrino appearance channel $P(\nu_{\mu}\rightarrow \nu_e)$ and the muon neutrino disappearance channel $P(\nu_{\mu} \rightarrow \nu_{\mu})$. Concretely, we will explore the complementary to matter NSI among future LBL experiments namely, ESSnuSB, T2HKK and a DUNE-like experiment. 

The structure of the paper is as follows. In Sec.~\ref{frame}, we
review and develop the framework of matter
NSI. Sec.~\ref{simulation} explains the characteristics and
assumptions made in our simulation. Sensitivities and main results are
developed in Secs.~\ref{sens} and \ref{res}. Finally, we give our
conclusions in Sec.~\ref{con}.

\section{Framework}
\label{frame}

Different kinds of physics beyond the Standard Model can be studied,
from the phenomenological point of view, using the formalism of
nonstandard neutrino interactions (NSI). Essentially, NSI describes
any new physics that, at low energies, can be parametrized by the
Lagrangian

\begin{equation}
    \label{eq:lnsi}
    \mathcal{L}= -2 \sqrt{2} G_{F} \epsilon_{\alpha \beta}^{fC}(\Bar{\nu}_{\alpha}\gamma^{\mu}P_{L}\nu_{\beta}) (\Bar{f}\gamma_{\mu}P_{C}f).
\end{equation}
Here, $\alpha, \beta = e, \mu, \tau$ refer to the neutrino flavor, $f
= e, u, d$ stand for the target fermions, $P$ indicates the projector
operator, with the superscript $C = L, R$ indicating the chirality of
the $ff$ current; finally, $\epsilon_{\alpha \beta}^{f C}$ are the
strengths of the NSI.

For neutrinos propagating through the Earth on a matter background,
the NSI contribution to the Hamiltonian will be proportional to the
corresponding fermion density ($e$, $u$, and $d$) times the given NSI
parameter. We can parametrize the three relevant contributions in a
single parameter, $\epsilon_{\alpha \beta}$, as
\begin{equation}
    \label{eq:epsdef}
    \epsilon_{\alpha \beta} = \sum_{\small{f  = e, u, d}} \epsilon_{\alpha \beta}^{f} \frac{N_{f}}{N_{e}}  ~: = \sum_{\small{f  = e, u, d}} (\epsilon_{\alpha \beta}^{f L} + \epsilon_{\alpha \beta}^{f R}) \frac{N_{f}}{N_{e}},
\end{equation}
where $N_{f}$ corresponds to the number density of the $f$ fermion. In
this article, we will work in the approximation where, for the Earth,
$N_{n} \simeq N_{p} = N_{e}$, then $N_{u} \simeq N_{d} \simeq 3
N_{e}$. Therefore,
\begin{equation}
    \epsilon_{\alpha \beta} \simeq \epsilon_{\alpha \beta}^e+ 3\epsilon_{\alpha \beta}^{u}+ 3\epsilon_{\alpha \beta}^{d}.
\end{equation}

With this notation, we can write the NSIs contribution to the
effective Hamiltonian of the neutrino propagation in matter, in the
flavor basis,\footnote{Notice that we use the Hermiticity condition of the interaction, that is, $\epsilon_{\alpha \beta}^{f C} = (\epsilon_{ \beta \alpha}^{f C})^{*}$.} as

\begin{equation}
\label{eq:hamil}
\begin{aligned}
H= \frac{1}{2 E}\Bigg[ U^{\dagger} M^2 U + A^{\text{CC}} \left(
\begin{array}{ccc}
 1 + \epsilon_{ee} & \epsilon_{e \mu} & \epsilon_{e \tau} \\
 \epsilon_{e \mu}^{*} & \epsilon_{\mu \mu} &  \epsilon_{ \mu \tau} \\
 \epsilon_{e \tau}^{*} & \epsilon_{ \mu \tau}^{*} & \epsilon_{ \tau \tau} \\
\end{array}
\right) \Bigg].
\end{aligned}
\end{equation} 
Here, $E$ is the neutrino energy, $U : = R_{23}(\theta_{23})
U_{13}(\theta_{13}, \delta) R_{12}(\theta_{12})$ is the leptonic
mixing matrix, $M^2=\text{diag}(0, \Delta m_{21}^2, \Delta m_{31}^2)$
is the diagonal mass-matrix, $A^{\text{CC}} = 2 \sqrt{2} G_{F} N_e E$
is the standard charged current matter potential. We consider complex
NSI, where $\epsilon_{\alpha \beta} = |\epsilon_{\alpha \beta}| e^{i
  \phi_{\alpha \beta}}$ for $\alpha \neq \beta$, which may contribute
to $CP$-violation in the leptonic sector.

Using this Hamiltonian, we compute the exact survival and conversion
probability expressions. This computation was done by using the GLoBES
software~\cite{Huber:2004ka,Huber:2007ji}, especially its additional
NSI tool~\cite{Kopp:2006wp,Kopp:2007rz}. For instance, approximate
analytic expressions for the oscillation probability in the presence
of matter NSI exist in the literature for both $P(\nu_{\mu}
\rightarrow \nu_{\mu})$
\cite{Kikuchi:2008vq,Liao:2016orc,Chatterjee:2021wac} and $P(\nu_{\mu}
\rightarrow \nu_e)$ channels \cite{Kikuchi:2008vq,
  Liao:2016hsa,Chatterjee:2020kkm, Capozzi:2019iqn}.

In this work we will explore the NSI as a solution to the discrepancy
observed between the central values of the NOvA and T2K leptonic-$CP$
measurements. According to the scenario investigated
in~\cite{Chatterjee:2020kkm,Denton:2020uda}, T2K (which is a
practically vacuum oscillation experiment), determines the true value
of the leptonic $CP$-phase $\delta_{CP} \sim 1.4
\pi$~\cite{T2K2020, T2K:2023smv}. On the other hand, the NOvA
experiment (with more matter interactions) measures a value of
$\delta_{CP}\sim 0.8 \pi$~\cite{Nova2020}, in presence of
matter NSI the $CP$-phase $\delta_{\text{NOvA}}
\sim\delta_{\text{T2K}}+\phi$, where the extra $CP$-violating phase $\phi$
is induced from the NSI effects, either $\phi = \{ \phi_{e \mu}~
\text{or} ~\phi_{e \tau} \} \sim 3/2 \pi$ and effective couplings
$|\epsilon_{e \mu}| \sim |\epsilon_{e \tau}| \sim
0.2$~\cite{Chatterjee:2020kkm,Denton:2020uda}. At the probability
level:
\begin{equation}
\label{posc}
    P (\epsilon = 0, \delta_{\text{meas}}) = P (\epsilon, \delta_{\text{true}}).
\end{equation}
The detailed explanation of the phase relationships can be found elsewhere (e.g., in the Supplemental Material of Ref.~\cite{Denton:2020uda}). This scenario, where ambiguities in the determination of the leptonic $CP$-violating phase $\delta_{CP}$
at NOvA and T2K arise via flavor changing matter NSI parameters $
(|\epsilon_{e \tau}|, \phi_{e \tau})$ was explored in
\cite{Forero:2016cmb}. A related study at the probability level can be
found in the literature~\cite{Flores:2018kwk}. Furthermore, as shown by the authors of~\cite{Bakhti:2016prn}, the combination of the proposed MOMENT experiment with the NOvA and T2K datasets can help in determining the Dirac $CP$-phase $\delta_{CP}$ in the presence of NSI.

For instance, relatively large NSI ($\epsilon_{e \tau} \sim 0.3-0.5$)
can appear in radiative neutrino mass models (see, e.g.,~\cite{Forero:2016ghr,Babu:2019mfe}). Moreover, sizable NSI can be
induced in models with light
mediators~\cite{Farzan:2015doa,Farzan:2015hkd, Farzan:2016wym, Denton:2018xmq, Dey:2018yht, Farzan:2019xor}. Alternatively, the
determination of the neutrino mass ordering might be spoiled by the
presence of matter NSI \cite{Capozzi:2019iqn}. Lately, the authors of
Ref.~\cite{Denton:2022pxt} investigate the potential of the Deep
Underground Neutrino Experiment (DUNE) to probe new physics scenarios,
which are motivated by the aforementioned NOvA and T2K results.

\section{Experimental setup and simulation}
\label{simulation}
This section presents the characteristics and assumptions performed in
our study. For this work, we will focus on the ESSnuSB proposal and
its impact in a combined analysis with a future DUNE-like
experiment. Moreover, we will analyze a two detector ESSnuSB
configuration as well as the T2HKK proposal. The European Spallation
Source plans to start operations in the year 2035. It will be the most
powerful spallation source in Europe, and a long-baseline neutrino
program is contemplated~\cite{Abele:2022iml}. This LBL proposal,
ESSnuSB, considers using an intense proton beam of $2.5$~GeV in Lund,
Sweden. Its main purpose will be the search for the neutrino $CP$-phase
by locating a far detector inside a mine. Among the options available
in the region, the more promising, according to their particle physics
program~\cite{Abele:2022iml}, are the ones located at $360$ and
$540$~km. However, other mines exist at different distances, such as
$260$ and $1090$~km. Besides searching for $CP$-violation, ESSnuSB also
plans to search for cosmological and supernovae neutrinos as well as
to set new limits to the proton lifetime.

In the case of a DUNE-like experiment, the proposal contemplates a
detection technology based on liquid argon, with a $40$~kton mass,
located at $1300$~km from the source, a proton beam with a $1.2$~MW
power.

On the other hand, the Tokai to Hyper-Kamiokande and Korea (T2HKK)
program~\cite{Hyper-Kamiokande:2016srs} consists of a two-detector
experimental setup for the discovery of the Dirac $CP$-violating phase
$\delta_{CP}$, which employs a (near) detector located at the
Kamioka site at 295 km from the beam at Japan Proton Accelerator
Research Complex (J-PARC) in Japan and the second (far) detector at a
distance of 1100 km in Korea. Several off-axis angles (OA$^{\circ}$)
fluxes are in consideration, from which we will use the two-degree
off-axis (OA2$^{\circ}$) configuration of the fluxes from both the
near and far detectors.

We use the GLoBES software~\cite{Huber:2004ka,Huber:2007ji} and its
additional NSI tool~\cite{Kopp:2006wp,Kopp:2007rz} in this
analysis. Regarding the ESSnuSB setup, the matter densities are
assumed to be $\rho = (2.6, 2.75, 2.8, 2.84)$ g/cm$^3$ for the (200,
360, 540, 1090) km baselines, respectively. In addition, for the case
of a single experimental arrangement with two baselines, we have
considered two identical detectors, an \emph{intermediate} detector at
the closest distance from the neutrino beam and a far detector at the
longest distance, with a total detector mass $m_{\text{tot}} =
m_{\text{Near}}+ m_{\text{Far}} = 538$ kt, evenly distributed among
the detectors. Moreover, we suppose the default systematic
uncertainties of $10\%~\text{and}~15\%$ normalization for signal and
background, respectively, and a $0.01\%$ energy calibration error for
all types of events. 
According to Ref.~\cite{Alekou:2022emd}, the effect of correlation for a bin-to-bin analysis is relatively mild. Therefore, as a crude approximation, we implement the systematic uncertainties as uncorrelated~\cite{Blennow:2015nxa,Ishitsuka:2005qi}. 
Other specifications for the ESSnuSB follow closely the previous work of 
Ref.~\cite{Cordero:2022fwb}.

For the DUNE-like experiment, we use the specifications and available
files from the Technical Design Report (TDR) \cite{DUNE:2020jqi,
  DUNE:2020ypp}, which uses a 120 GeV proton beam of 1.2 MW power,
with a matter density assumed to be $\rho = 2.848$ g/cm$^3$. Moreover,
this configuration considers a 40 kt detector with an exposure of 13
years equally divided among (anti)neutrino modes.

Regarding the T2HKK simulation, the fluxes, cross sections, and
efficiencies follow the description from Ref.~\cite{T2K:2001wmr},
corresponding to a $2^{\circ}$ off-axis (OA$2^{\circ}$) flux
configuration at both the near and far detectors. Furthermore, the
energy resolution has a width $\sigma_{E}/E$ of $8.5 \%$ for both
$e^-$ and $\mu^{-}$ respectively. Moreover, the near detector is
located at 295 km, with a second identical detector at 1100 km. The
total mass is 560 kt, equally divided among the detectors
\cite{Fukasawa:2016lew}. The total exposure time corresponds to 13
MW$\cdot$yr, which corresponds to a total $2.7 \times 10^{22}$ POT, 10
years of running time at 1.3 MW power \cite{Liao:2016orc} evenly
distributed among (anti)neutrino modes. In addition, a matter density
of $\rho = 2.6~\text{g/cm}^3$ at the 295 km and $\rho =
2.84~\text{g/cm}^3$ at the 1100 km baseline has been considered.  We
assume a normalization uncertainty of 2.5$\%$ for the signal rates and
5$\%$ (20$\%$) for the appearance (disappearance) background rates
\cite{Liao:2016orc, Ishitsuka:2005qi}. As in the case of ESSnuSB, we implement the systematic uncertainties as uncorrelated, an approach that has been followed by other authors~\cite{Ghosh:2017ged}.

As far as neutrino oscillation parameters are concerned, the true
values used in this analysis are: $\Delta m^{2}_{21}=7.5 \times
10^{-5}~\text{eV}^{2}$, $\Delta m^{2}_{31} = 2.55 \times
10^{-3}~\text{eV}^2$, $\theta_{12} = 34.3^{\circ}$, $\theta_{13} =
8.53^{\circ}$, $\theta_{23} = 49.26^{\circ}$, $\delta_{CP} =1.4
\pi$; corresponding to the best-fit values for normal ordering (NO)
from Salas \emph{et al.}~\cite{deSalas:2020pgw} (except for
$\delta_{CP}$, unless otherwise specified, we take the best fit
from T2K~\cite{T2K2020,T2K:2023smv}), in all our results the NO is
considered.\footnote{We have verified that our results do not
  significantly change by using the best fit values from
  Ref.~\cite{Esteban:2020cvm}.} For oscillation parameter priors, we
assume a 1$\sigma$ error of $ 5\%$ for $\Delta m^{2}_{21}$, $\Delta
m^{2}_{31}$, $\theta_{12}$, and $\theta_{23}$. We also assume $3\%$
for $\theta_{13}$ and $ 10\%$ for the leptonic $CP$-violating phase
$\delta_{CP}$~\cite{ESSnuSB:2013dql}. Furthermore, for all the
baselines in consideration, a 1$\sigma$ uncertainty of $3\%$ on the
standard matter density was assumed. Besides, when we consider NSI
matter effects from the ($e-\mu$) sector, we set $|\epsilon_{e\mu}|=
0.19$ and $\phi_{e \mu}= 1.5\pi$ consistent with their best fit from
Ref.~\cite{Denton:2020uda}. In addition, matter NSI effects from the
($e-\tau$) sector are set to their best-fit values from
Ref.~\cite{Chatterjee:2020kkm}, $|\epsilon_{e\tau}|= 0.275$ and
$\phi_{e \tau}= 1.62\pi$, all the remaining matter NSI parameters were
fixed to zero.

\section{NSI Sensitivity}

\label{sens}

In this section, we outline the calculation of sensitivities to the
NSI parameters. We employ a chi-squared test to quantify the
statistical significance of matter NSI oscillations using neutrino and
antineutrino datasets. The $\chi^2$ function\footnote{More details
  on the implementation of the $\chi^2$ function, systematical errors
  and priors in the GLoBES software~\cite{Huber:2004ka,Huber:2007ji}
  can be found in \cite{Huber:2002mx}.} is given as
\begin{equation}
    \chi^2 = \sum_{\ell} \tilde{\chi}^2_{\ell} +\chi^2_{\text{prior}},
\end{equation}
where the corresponding $\tilde{\chi}^2_{\ell}$ function for each channel $\ell= \big( \nu_{\mu}(\Bar{\nu}_{\mu})\rightarrow \nu_{e} (\Bar{\nu}_{e}),~ \nu_{\mu}(\Bar{\nu}_{\mu})\rightarrow \nu_{\mu} (\Bar{\nu}_{\mu}) \big)$, which in the large data size limit is 
\begin{equation}
    \tilde{\chi}^2_{\ell} = \min_{\xi_{j}} \Bigg[  \sum_{e}^{N, F}\sum_{i}^{n_{\text{bins}}} \frac{\big(N_{i, e}^{3\nu}-N_{i, e}^{3 \nu+\text{NSI}}( \Omega, \Theta, \{\xi_{j}\}) \big)^2}{\sigma_{i, e}^2} +\sum_{j}^{n_{\text{syst.}}} \Big(\frac{\xi_{j}}{\sigma_{j}} \Big)^2  \Bigg].
\end{equation}
The $N_{i}^{3\nu}$ are the simulated events at the $i$th energy bin
considering the standard three neutrino oscillations
framework. $N_{i}^{3\nu +\text{NSI}}$ are the computed events at the
$i$th energy bin with the model assuming matter NSI
oscillations. $\Omega = \{\rho, \theta_{12}, \theta_{13}, \theta_{23},
\delta_{CP}, \Delta m_{21}^2, \Delta m^2_{31}\}$ is the set of
matter density and oscillation parameters, $\Theta =
\{|\epsilon_{e\mu}|, \phi_{e \mu}, |\epsilon_{e\tau}|, \phi_{e
  \tau}\}$ is the set of NSI parameters and $\{\xi_{j}\}$ are the
nuisance parameters to account for the signal, background
normalization, and energy calibration systematics
respectively. Moreover, $\sigma_{i} = \sqrt{N_{i}^{3\nu}}$ is the
statistical error in each energy bin, while $\sigma_{j}$ are the
signal, background normalization, and energy calibration errors (see
Sec.~\ref{simulation}). The summation index $e$ runs over either the
corresponding (single) experiment at two different baselines (near and
far) or the combination of two separate experimental
setups. Furthermore, the implementation of external input for the
standard oscillation parameters on the $\chi^2$ function is performed
via Gaussian priors
\begin{equation}
    \chi^2_{\text{prior}}= \sum_{k}^{n_{\text{priors}}} \Bigg\{  \frac{\big(\Omega_{k,\text{true}}-\Omega_{k,\text{test}}\big)^2}{\sigma^2_{k}}+\frac{\big(\Theta_{k,\text{true}}-\Theta_{k,\text{test}}\big)^2}{\sigma^2_{k}} \Bigg\}, 
\end{equation}
the central values of the oscillation parameter priors $\Omega_{k}$
are set to their true or best-fit value for normal ordering
\cite{deSalas:2020pgw}, and the central values of the matter density
change for the different experiments in consideration (see
Sec.~\ref{simulation}). $\sigma_{k}$ is the uncertainty on the
oscillation prior, which corresponds to a 1$\sigma$ error of $5\%$ for
$\Delta m^{2}_{21}$, $\Delta m^{2}_{31}$, $\theta_{12}$, and
$\theta_{23}$, $3\%$ for $\theta_{13}$, $ 10\%$ for the leptonic
$CP$-violating phase $\delta_{CP}$~\cite{ESSnuSB:2013dql} and
$3\%$ for the matter density $\rho$. Furthermore, the central values
of the priors $\Theta_{k}$ change depending on the hypothesis in
consideration. When marginalization over NSI parameters from either the $e-\mu$ or $e-\tau$ sector is required, a
1$\sigma$ error $\sigma_{k}$ of $30\%$ is assumed. The summation index
$k$ runs over the corresponding test oscillation parameters to be
marginalized. Moreover, the expected number of events at the $i$th
energy bin was calculated as in~\cite{Huber:2002mx} (see,
e.g., Ref.~\cite{Cordero:2022fwb} as well).

\section{Results}
\label{res}

In this section, we present our results for the different experimental
configurations, emphasizing the matter NSI scenario (considering one
$\epsilon_{\alpha \beta}$ parameter at a time) motivated by the
discrepancy in the measurement of the leptonic $CP$-violating phase
$\delta_{CP}$ experienced by NOvA and
T2K~\cite{Chatterjee:2020kkm, Denton:2020uda}. We will focus on the
constraints on the $\delta_{CP}$ phase in the presence of
NSI. We will show the complementary between the ESSnuSB and other
current proposals such as T2HKK and, especially, DUNE.

We will show the sensitivity to matter NSI parameters by combining two
different long baseline experiments. We have studied the case of the
ESSnuSB and its combined restrictive power when we also consider a
DUNE-like (TDR) experiment. In this case we have taken into account
either the ESSnuSB setup at 360 km or 540 km
\cite{ESSnuSB:2021azq}. Since the ESSnuSB is still in a proposal, we
have also computed the results of a ESSnuSB configuration with two LBL
detectors, to see if the combination may help to improve the
robustness of their $CP$-phase measurement. We have also confronted this
result with the case of the T2HKK proposal, that already considers two
detectors in its experimental setup.
 
To illustrate, our analysis we will contrast two different cases: the
standard three neutrino oscillation scenario and the case where NSI is
present in matter evolution. In the first case, we calculate our
results assuming the standard oscillation picture and fitting the data
assuming a 
given true value of $\delta_{CP}$ and
$\sin^2(\theta_{23})$ and varying for the test values. As stated in Sec. \ref{simulation}, we consider as true values $\theta_{23} = 49.26^{\circ}$ and $\delta_{CP}= 1.4\pi$. For the standard + NSI case, we have considered
either the NSI parameters $\phi_{e\mu}$ and $|\epsilon_{e \mu}|$ or
$\phi_{e\tau}$ and $|\epsilon_{e \tau}|$ as free parameters that are
marginalized away in the fit, again for the same fixed true values of
$\delta_{CP}$ and $\sin^2(\theta_{23})$. We will consider NSI parameters that are different from the SM case ({\it i.e.} different from zero), but that mimic the SM solution in the matter case as noted by the authors of Refs.~\cite{Chatterjee:2020kkm,Denton:2020uda}. In the following, we
will show the results of the standard-only analysis as solid lines,
while we will show the NSI case with dashed lines. Moreover, we display in green solid contours the expected allowed regions considering all the NSI entries (SM+Full NSI), marginalizing over all NSI parameters, taking into account the bounds from IceCube and global analysis with and without CE$\nu$NS data from COHERENT~\cite{Esteban:2018ppq, Ice2019}. \footnote{As shown in Figs.~9 and 12 of \cite{Esteban:2018ppq}, the inclusion of CE$\nu$NS data (COHERENT) does not improve the constraints on the NSI parameters for $\Delta \chi^2 \lesssim 2\sigma$. Besides, the incorporation of COHERENT constraints are valid under certain considerations \cite{Denton:2020uda, Esteban:2018ppq}.} The diagonal NSI parameters were varied from $|\epsilon_{ee}-\epsilon_{\mu\mu}|\leq 0.5$, $|\epsilon_{\tau \tau}-\epsilon_{\mu\mu}|\leq 0.04$, and we have considered $\epsilon_{\mu\mu}=0$ taking advantage of the freedom to redefine the diagonal elements up to a global constant~\cite{deGouvea:2015ndi,Bakhti:2016prn}. The corresponding off-diagonal entries were varied from $|\epsilon_{e \mu}| \leq 0.19$, $|\epsilon_{e \tau}| \leq 0.2$, $|\epsilon_{ \mu \tau}| \leq 0.023$, and, finally, the extra NSI $CP$-phases were varied from $ 0\leq \phi_{\alpha \beta} \leq 2 \pi$. 
Although we do not consider the LMA-Dark region~\cite{Miranda:2004nb} in the SM+Full NSI analysis, we have also studied it and found that in this case the DUNE sensitivity to the $CP$-violating phase can be completely lost. For all cases, all the remaining standard oscillation parameters were marginalized.

\subsection{DUNE + ESSnuSB}
An interesting case is the expected sensitivity from the combination
of a future DUNE-like detector and ESSnuSB with a baseline of 540
km. Before showing our results, we briefly summarize the main
complementary characteristics of these two projects.

For the corresponding LBL experiments of interest, we can estimate an
average matter density $\rho \sim $ 3.0 g/cm$^3$. Besides, the
approximate energy for MSW resonance occurs at $E\sim$ GeV. Since the
flux for ESSnuSB (360 km or 540 km) peaks at $E \sim \mathcal{O}(0.1)$
GeV (see Fig.~1 of Ref.~\cite{Blennow:2019bvl}), matter effects for
this facility are not expected to be significant. Therefore, for our
purposes, in the case of a standard three neutrino oscillation
framework (SM) as well as the SM$+$NSI case, we assume
$\delta_{CP}\sim 1.4 \pi$ for ESSnuSB, which is consistent
with the best fit from the similar experiment T2K~\cite{T2K2020,T2K:2023smv}. Also, within this scenario, ESSnuSB should be able to measure a $\delta_{CP}$ value that is unaffected by the presence of matter NSI.

For the case of DUNE, with an average neutrino energy $E\sim 3$~GeV,
the situation is opposite, and matter effects are expected to be
relevant. In this case, we will consider $\delta_{CP} = 1.4\pi$ as the true value in both, the SM and SM$+$NSI scenarios. 

\begin{figure}[H]
		\begin{subfigure}[h]{0.47\textwidth}
			\caption{  }
			\label{ff1}
\includegraphics[width=\textwidth]{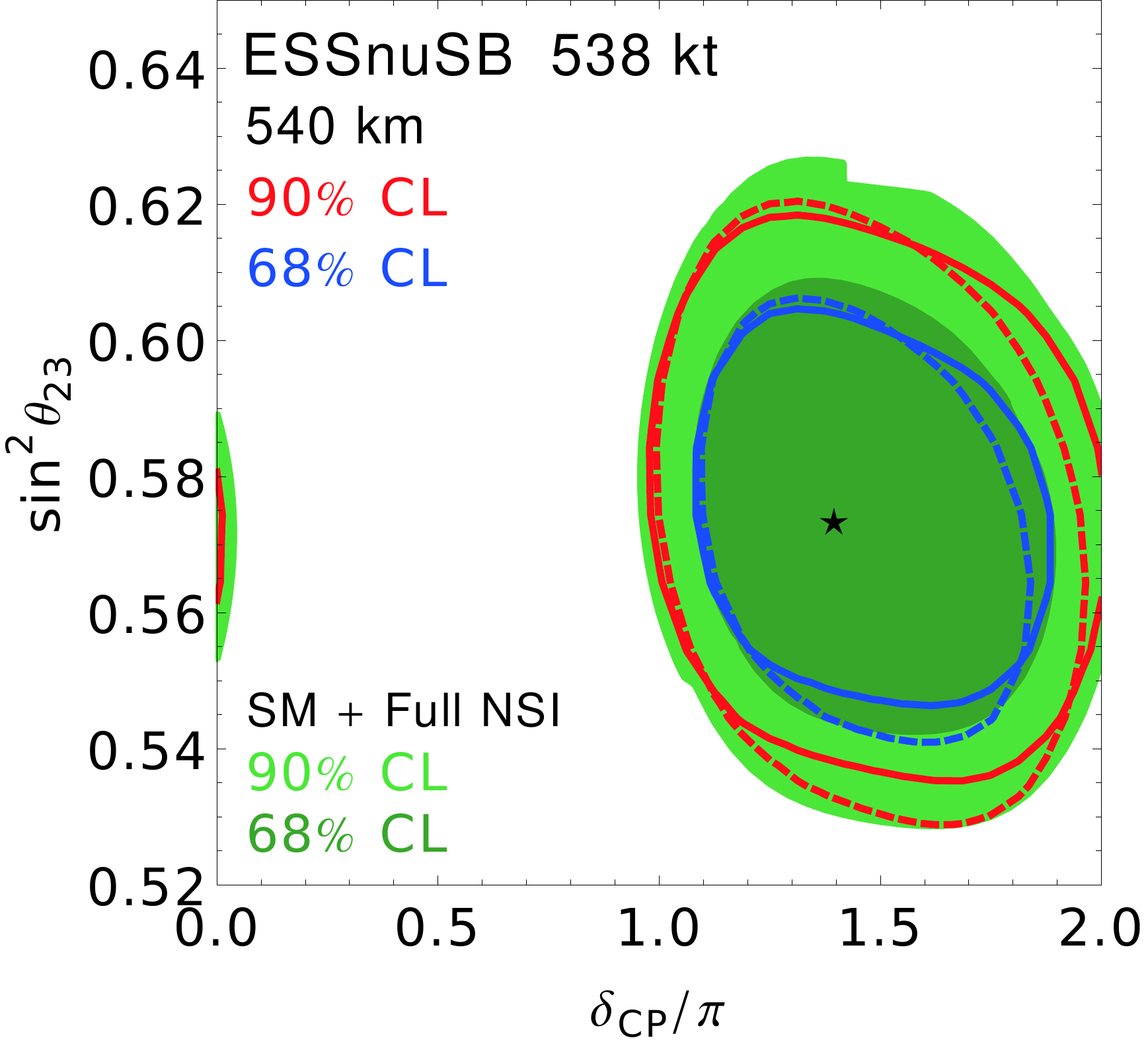}
		\end{subfigure}
		\hfill
		\begin{subfigure}[h]{0.47 \textwidth}
			\caption{}
			\label{ff2}
			\includegraphics[width=\textwidth]{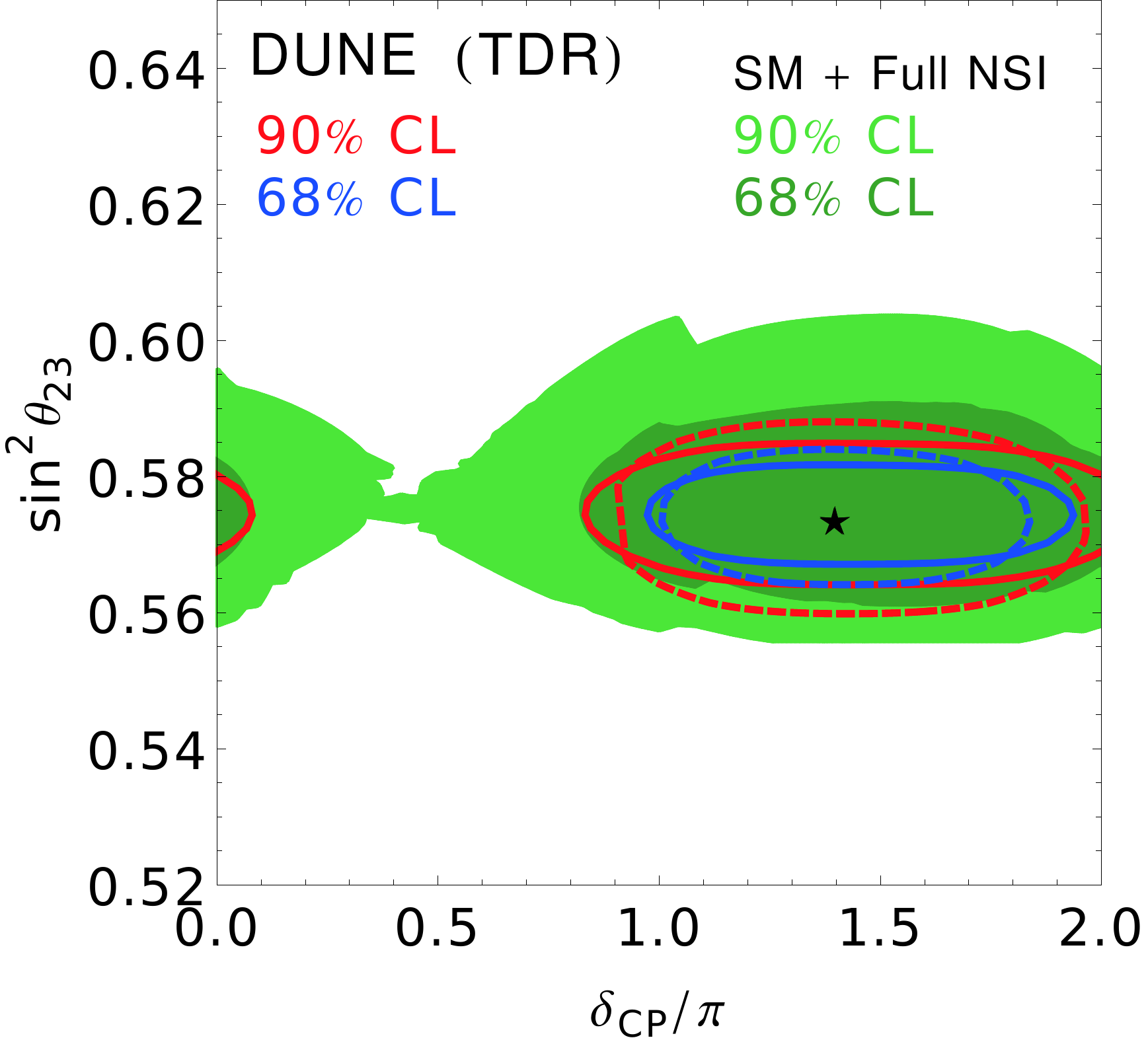}
		\end{subfigure}
		\hfill	
 \caption{Expected allowed regions in the ($\delta_{CP}$, $\sin^2 \theta_{23}$) plane. The standard 3$\nu$ scenario (SM) is displayed in (solid lines) while (dashed lines) show the case with (SM + NSI) assuming the best fit values ($|\epsilon_{e \mu}| =0.19$, $\phi_{e \mu}=1.5 \pi$) from Ref.~\cite{Denton:2020uda}. The left panel presents an ESSnuSB setup at 540 km from the source while the right panel sets the DUNE-like (TDR) configuration. Finally, we display in green solid contours the expected allowed regions considering all the NSI entries (SM+Full NSI), see text for a detailed explanation. 
  }
  \label{fdune}
\end{figure}

We illustrate the above discussion in Fig.~\ref{fdune}, where we show our results for the sensitivity to ($\delta_{CP}$, $\sin^2 \theta_{23}$) at 68$\%$ C.L. and 90$\%$ C.L. with and without the presence of NSI. In the left panel of this figure, we show the ESSnuSB 540 km configuration, and the corresponding case for a experiment of the type of DUNE appears in the right panel. As mentioned, the solid lines refer to the standard oscillation case, and the dashed ones stand for the NSI sensitivity. 
For both standard and NSI analysis, we have marginalized all other standard oscillation parameters. 
In the NSI case, we consider the flavor-changing ${e \mu}$ case, marginalized considering a 1$\sigma$ error of $30\%$ around their best fit ($|\epsilon_{e \mu}| =0.19$, $\phi_{e \mu}=1.5 \pi$) from~\cite{Denton:2020uda}. We can see that ESSnuSB sensitivity to the mixing angle is worse than the expected sensitivity in DUNE, while the sensitivity to the $\delta_{CP}$ phase is better for the ESSnuSB proposal. In addition, from the left panel of Fig.~\ref{fdune}, we observe that after matter NSI affects are included at ESSnuSB, the determination of the $\delta_{CP}$-phase and the mixing angle $\theta_{23}$ is practically unchanged.

\begin{figure}[H]
\includegraphics[scale=0.47]{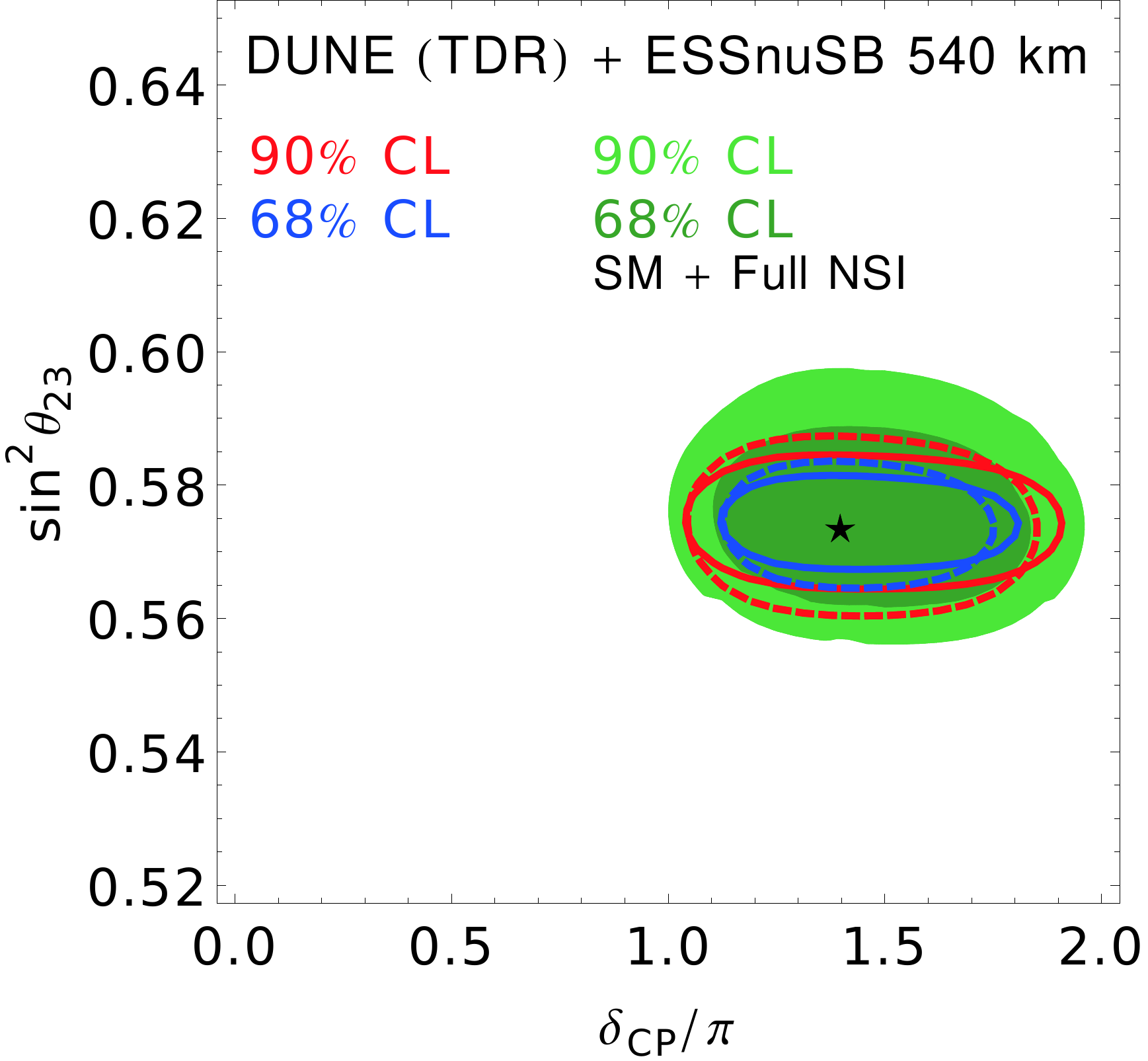}
		 \caption{Expected allowed regions in the ($\delta_{CP}$, $\sin^2 \theta_{23}$) plane plane for the combined DUNE-like (TDR) and ESSnuSB 540 km configuration. The standard 3$\nu$ oscillation framework (SM) is show in (solid lines) while (dashed lines) display the scenario with (SM + NSI) assuming the best fit values ($|\epsilon_{e \mu}|=0.19$, $\phi_{e \mu}=1.5 \pi$) from Ref.~\cite{Denton:2020uda}. Finally, we display in green solid contours the expected allowed regions considering all the NSI entries (SM+Full NSI), see text for a detailed explanation. }
  \label{fcz}
\end{figure}

In order to observe the effect of the combined sensitivity of these two proposals, we display in Fig.~\ref{fcz} our results of the allowed 68$\%$ C.L. and 90$\%$ C.L. contours in the ($\delta_{CP}$, $\sin^2 \theta_{23}$) plane for the combined DUNE-like and ESSnuSB 540 km setup. We have marginalized all other standard oscillation parameters.

\begin{figure}[H]
		\begin{subfigure}[h]{0.47\textwidth}
			\caption{  }
			\label{fk12}
\includegraphics[width=\textwidth]{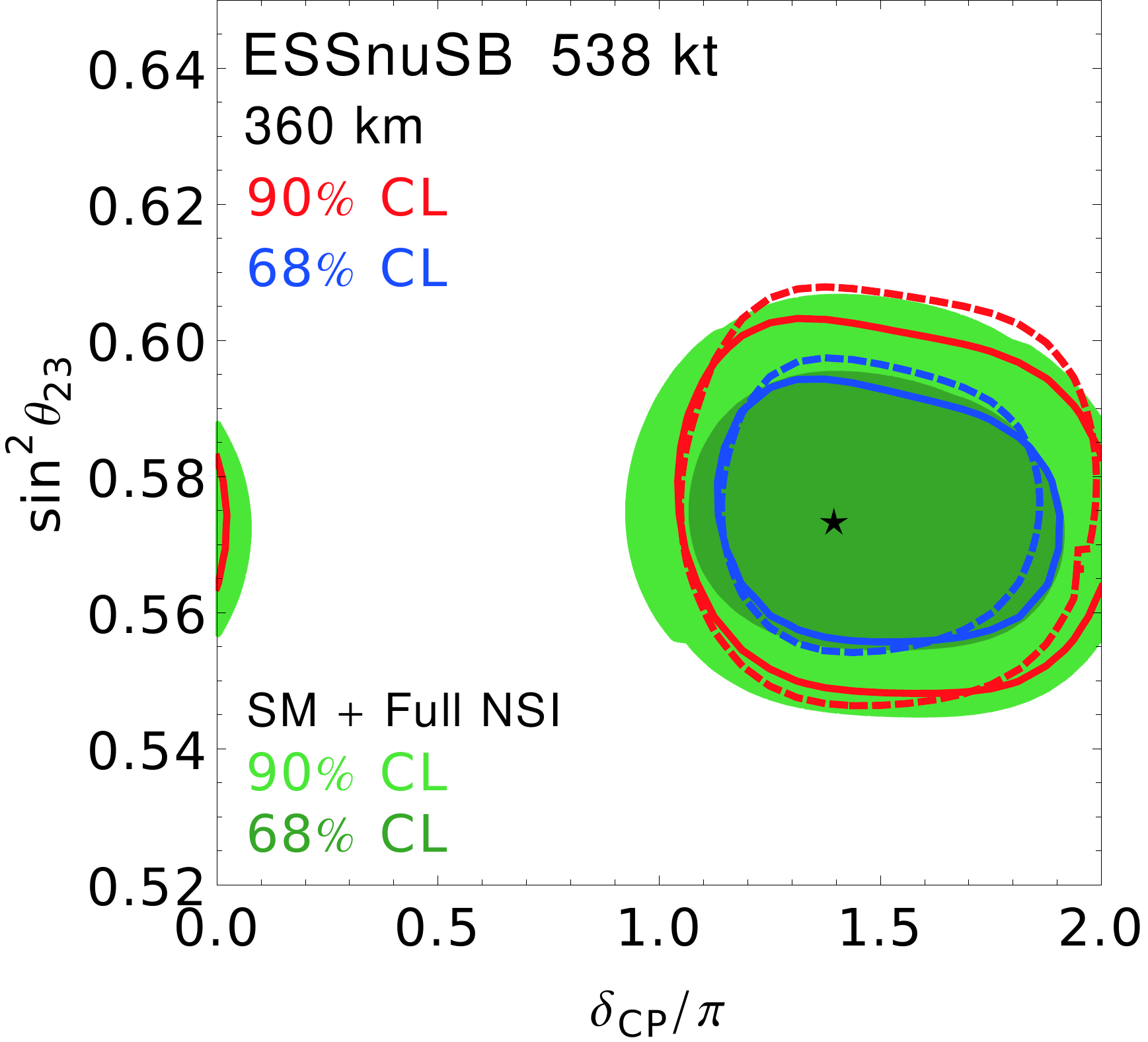}
		\end{subfigure}
		\hfill
		\begin{subfigure}[h]{0.47 \textwidth}
			\caption{}
			\label{fk13}
			\includegraphics[width=\textwidth]{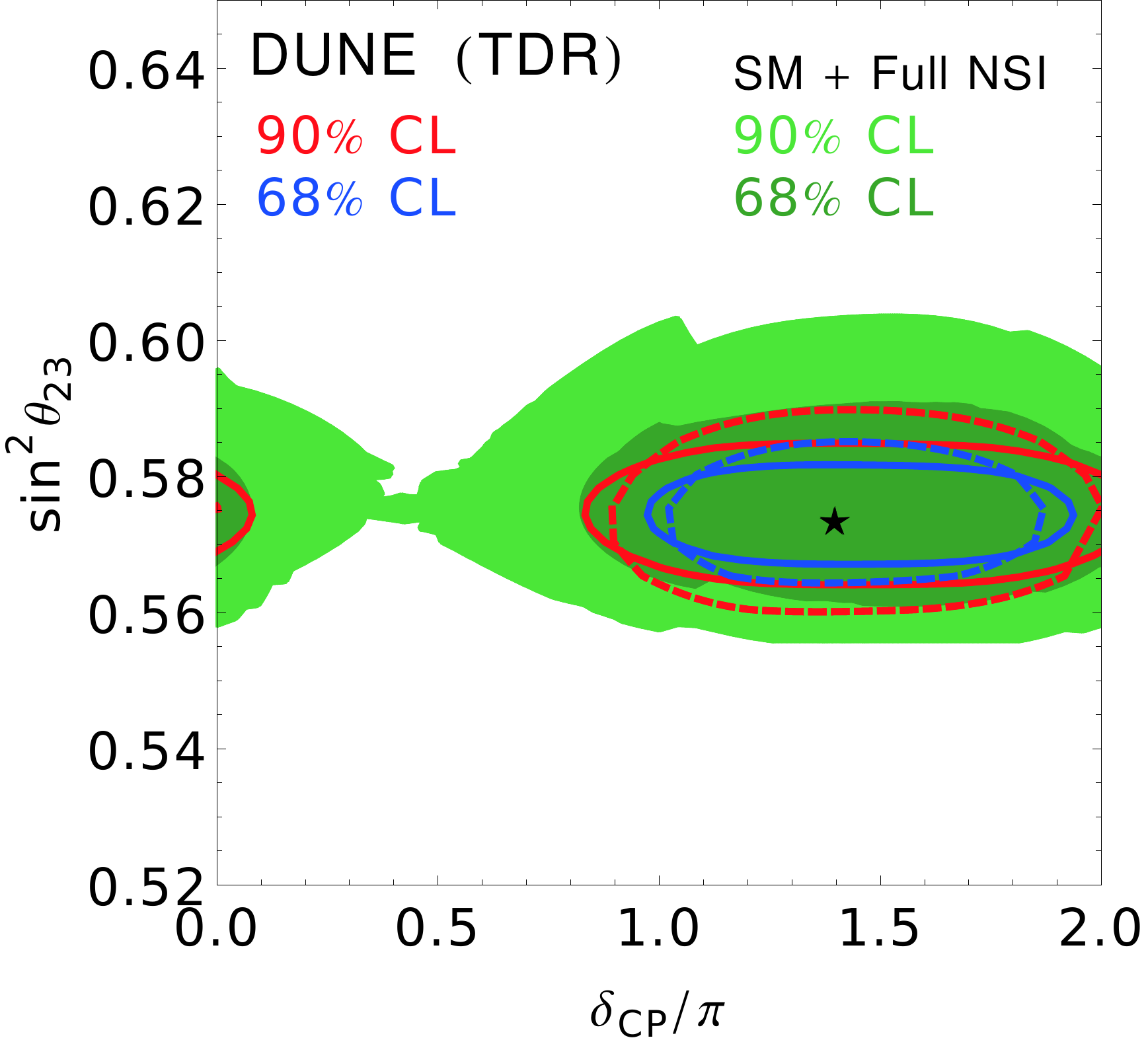}
		\end{subfigure}
		\hfill	
 \caption{Expected allowed regions in the ($\delta_{CP}$, $\sin^2 \theta_{23}$) plane. The standard 3$\nu$ framework (SM) is show in (solid lines) while (dashed lines) display the scenario with (SM + NSI) assuming the best fit values ($|\epsilon_{e \tau}| = 0.275$, $\phi_{e \tau} = 1.62 \pi$) from Ref.~\cite{Chatterjee:2020kkm}. The left panel presents an ESSnuSB setup at 360 km from the source while the right panel sets the DUNE-like (TDR) configuration. Finally, we display in green solid contours the expected allowed regions considering all the NSI entries (SM+Full NSI), see text for a detailed explanation. 
  }
  \label{figk}
\end{figure}

We can also study the case of an $e\tau$ flavor-changing NSI to
identify its impact on the sensitivity to the $CP$-violating phase. For
this purpose, in Fig.~\ref{figk}, we show the effect of nonzero
matter NSI parameters ($\epsilon_{e \tau}$) on the sensitivity to the
Dirac $CP$-violating phase $\delta_{CP}$ and the mixing angle
$\theta_{23}$. The corresponding allowed 68$\%$ C.L. and 90$\%$
C.L. contours in the ($\delta_{CP}$, $\sin^2 \theta_{23}$)
plane are displayed. The left panel refers to the ESSnuSB 360 km
configuration, while the right panel assumes the DUNE (TDR) case. We
have marginalized all other standard oscillation parameters. Besides,
the ($|\epsilon_{e\tau}|$, $\phi_{e\tau}$) parameters are
marginalized, considering a 1$\sigma$ error of $30\%$ around the best
fit ($|\epsilon_{e \tau}| =0.275$, $\phi_{e \tau}=1.62 \pi$) quoted
in~\cite{Chatterjee:2020kkm}. We observe that the ESSnuSB sensitivity
to the mixing angle is worse than the expected sensitivity in DUNE,
while the sensitivity to the $\delta_{CP}$ phase is slightly
better for the ESSnuSB proposal. Moreover, from Figs.~(\ref{fdune}) and (\ref{figk}), we observe that if $\delta_{CP}\sim 1.4 \pi$ is realized in nature, a DUNE-like experiment still allows the best fit from NOvA, $\delta_{CP}\sim 0.8 \pi$ at 90\% C.L. On the other hand, either of the ESSnuSB configurations will be able to exclude the NOvA best fit at 90\% C.L.

\begin{figure}[H]
		\includegraphics[scale=0.47]
			{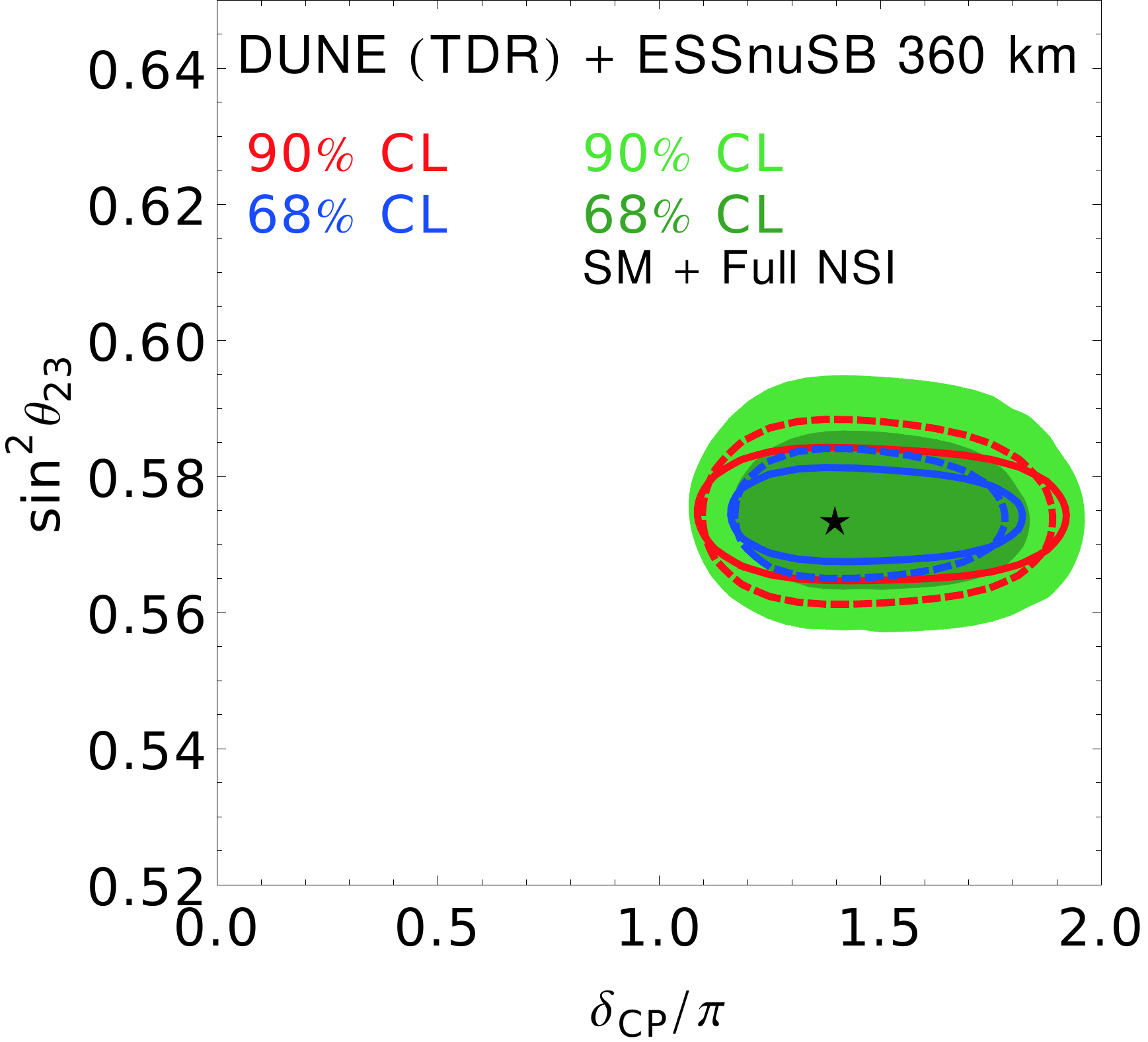}
			\caption{Expected allowed regions in the ($\delta_{CP}$, $\sin^2 \theta_{23}$) plane for the combined DUNE-like (TDR) and ESSnuSB 360 km configuration. The standard 3$\nu$ framework (SM) is show in (solid lines) while (dashed lines) display the scenario with (SM + NSI) where we set ($|\epsilon_{e \tau}| = 0.275$, $\phi_{e \tau} = 1.62 \pi$), corresponding to the best fit values from Ref.~\cite{Chatterjee:2020kkm}. Finally, we display in green solid contours the expected allowed regions considering all the NSI entries (SM+Full NSI), see text for a detailed explanation. 
  }
  \label{figz}
\end{figure}

We also show, in Fig.~\ref{figz}, the impact of nonzero matter NSI
parameters ($\epsilon_{e \tau}$) on the expected sensitivity of the
Dirac $CP$-violating phase $\delta_{CP}$ and the mixing angle
$\theta_{23}$. The corresponding expected sensitivity at 68$\%$
C.L. and 90$\%$ C.L. is displayed in this figure. We have marginalized
all other standard oscillation parameters.

\subsection{Two baseline configuration}

In this subsection, we investigate the constraining power to matter
NSI parameters using a single experiment with two baselines (ESSnuSB
or T2HKK). Such a setup can probe matter NSI effects within the NSI
framework~\cite{Chatterjee:2020kkm,Denton:2020uda}. Although the
ESSnuSB proposal has not considered the use of two different baselines
simultaneously, it is worth it to study this possibility now that the
project is in its first stages. In the following subsections, we will
assume that such a setup is feasible. We will work under the
hypothesis that both detectors can be considered as aligned with the
beam, implying an on-axis neutrino flux for both.

Regarding a two baseline configuration at ESSnuSB, from the existing
mines in Sweden, the corresponding Renstrom mine is located at a
distance of $L \sim$ 1090 km from the source at Lund, while the
Garpenberg mine is located at $L \sim$ 540 km. Both mines are at
roughly 1 km depth. Besides increasing sensitivity to
$\delta_{CP},~\theta_{23}$, and NSI from more exposure to SB
neutrinos, a second detector at the Renstrom mine will contribute to
the full physics program at ESSnuSB, which is the measurement of
proton decay, atmospheric (solar) neutrinos, supernovae neutrinos, and
geoneutrinos~\cite{Abele:2022iml, ESSnuSB:2013dql, Alekou:2022emd}.

Moreover, we consider as a possibility, a detector located at 200 km from
the source at Lund with a second detector placed at the Garpenberg
mine at 540 km. For instance, the authors of
Ref.~\cite{Chatterjee:2021xyu} explored the physics potential at
ESSnuSB (200, 360, 540) km baselines, respectively, within the
nonunitarity of the leptonic mixing matrix scenario. Although, no
available mines aligned with the Garpenberg mine, and the source at
Lund exists at 200 km. We will illustrate the benefits of such an
arrangement to probe the aforementioned matter NSI framework. However,
the ESSnuSB 200$-$540 km configuration might not accomplish the complete
ESSnuSB physics program~\cite{Abele:2022iml}.~\footnote{While it is true that the alignment of the two detectors setup is a rough approximation (there should be some off-axis neutrino-flux), our implementation of a two-detector ESSnuSB configuration gives a perspective on the future determination of the $CP$-phase in the presence of NSI at ESS, which can be complementary to the T2HKK proposal and may justify a further, more detailed study.}

We also consider the case of the T2HKK experiment, where we employ the
configurations from \cite{T2K:2001wmr}. This study can provide a
preliminary perspective of the T2HKK constraining power within the
matter NSI framework \cite{Chatterjee:2020kkm, Denton:2020uda}. For
this experimental setup, we also show the results for the expected
sensitivity to the flavor-changing NSI parameters for both
($\phi_{e\mu}$ and $|\epsilon_{e \mu}|$) and ($\phi_{e\tau}$ and
$|\epsilon_{e \tau}|$) since the perspectives are promising in this
case.

\begin{figure}[H]
		\begin{subfigure}[h]{0.49 \textwidth}
			\caption{  }
			\label{figuurr1}
\includegraphics[width=\textwidth]
                {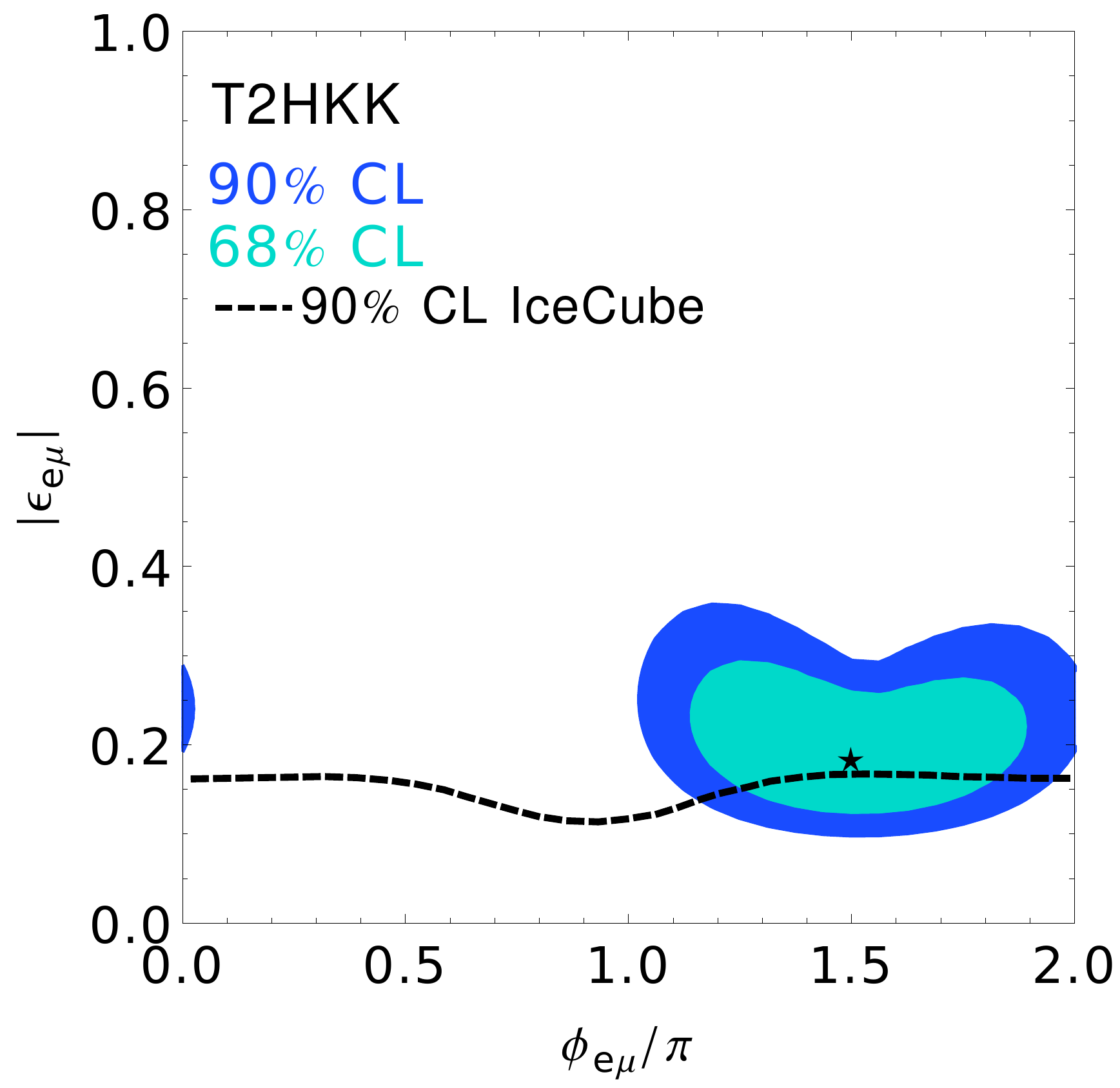}
		\end{subfigure}
		\hfill
		\begin{subfigure}[h]{0.49 \textwidth}
			\caption{}
			\label{figuurr2}
			\includegraphics[width=\textwidth]
               {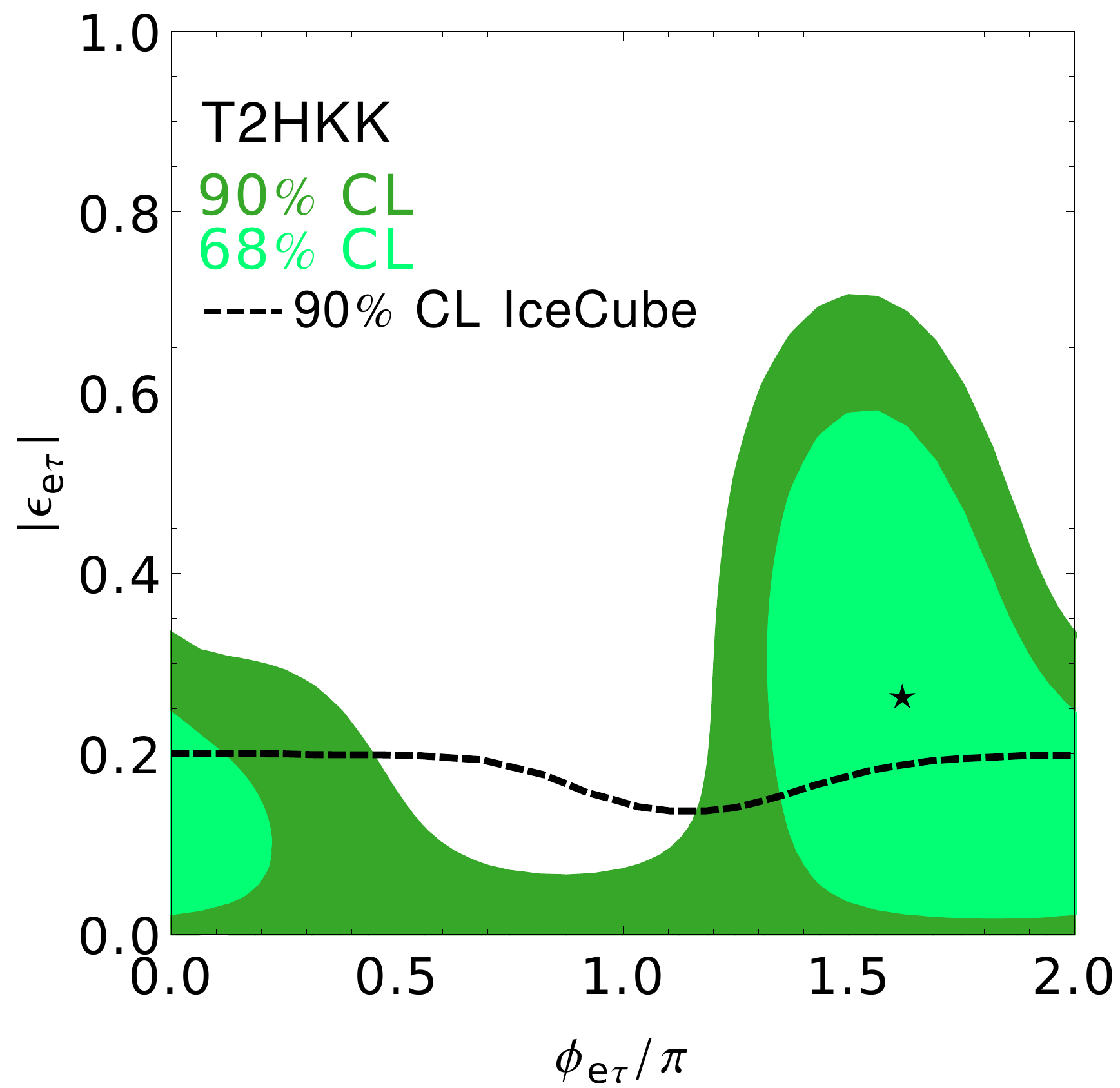}
		\end{subfigure}
		\hfill	
 \caption{Expected allowed regions in the ($|\epsilon|$, $\phi$) plane for the T2HKK OA2$^{\circ}$ configuration. Assuming NSI best fit values from Refs.~\cite{Chatterjee:2020kkm, Denton:2020uda}. Furthermore, we also show the 90$\%$ C.L. bounds from the IceCube DeepCore data~\cite{Ice2019}. }
  \label{figr}
\end{figure}

As shown in Fig.~\ref{figr}, the expected allowed contours in the
corresponding ($|\epsilon|$, $\phi$) planes are consistent with the
allowed regions determined by the combination of the NOvA and T2K data
sets from the left panel of Fig.~2
\cite{Chatterjee:2020kkm}. However, as already noticed, the IceCube
DeepCore~\cite{Ice2019} constraints exclude most part of the preferred
parameter space. For comparison, we can notice that, as discussed in
Ref.~\cite{Denton:2022pxt}, a DUNE-like experiment will be able to
determine the matter NSI parameters from the ($e-\mu$) and ($e-\tau$)
sectors with a precision of around [10$-$20$]\%$ for the NO.

\subsubsection{Impact of NSI on oscillation precision measurements}

In this subsection, we consider the effects of matter flavor changing
($e-\mu$) NSI parameters on the oscillation parameters. More
precisely, the expected allowed regions in the
($\delta_{CP},~\sin^2 \theta_{23}$) plane for both, the SM as
well as the (SM + NSI) scenario will be shown. Both electron neutrino
appearance and muon neutrino disappearance events are considered in
our analysis, while all the remaining oscillation parameters are
marginalized. Furthermore, for the matter NSI, the
($|\epsilon_{e\mu}|$, $\phi_{e\mu}$) parameters are marginalized,
considering a 1$\sigma$ error of $30\%$ around the true values
$(|\epsilon_{e\mu}|=0.19, \phi_{e\mu}=1.5 \pi)$, which were fixed to
their best fit from Ref.~\cite{Denton:2020uda} (see, e.g.,
Fig.~\ref{fcz}). In addition, we display in solid green contours (SM+Full NSI) the expected sensitivities marginalizing over all the matter NSI parameters.

\begin{figure}[H]
		\begin{subfigure}[h]{0.33 \textwidth}
			\caption{  }
			\label{fb12a}
\includegraphics[width=\textwidth]
		{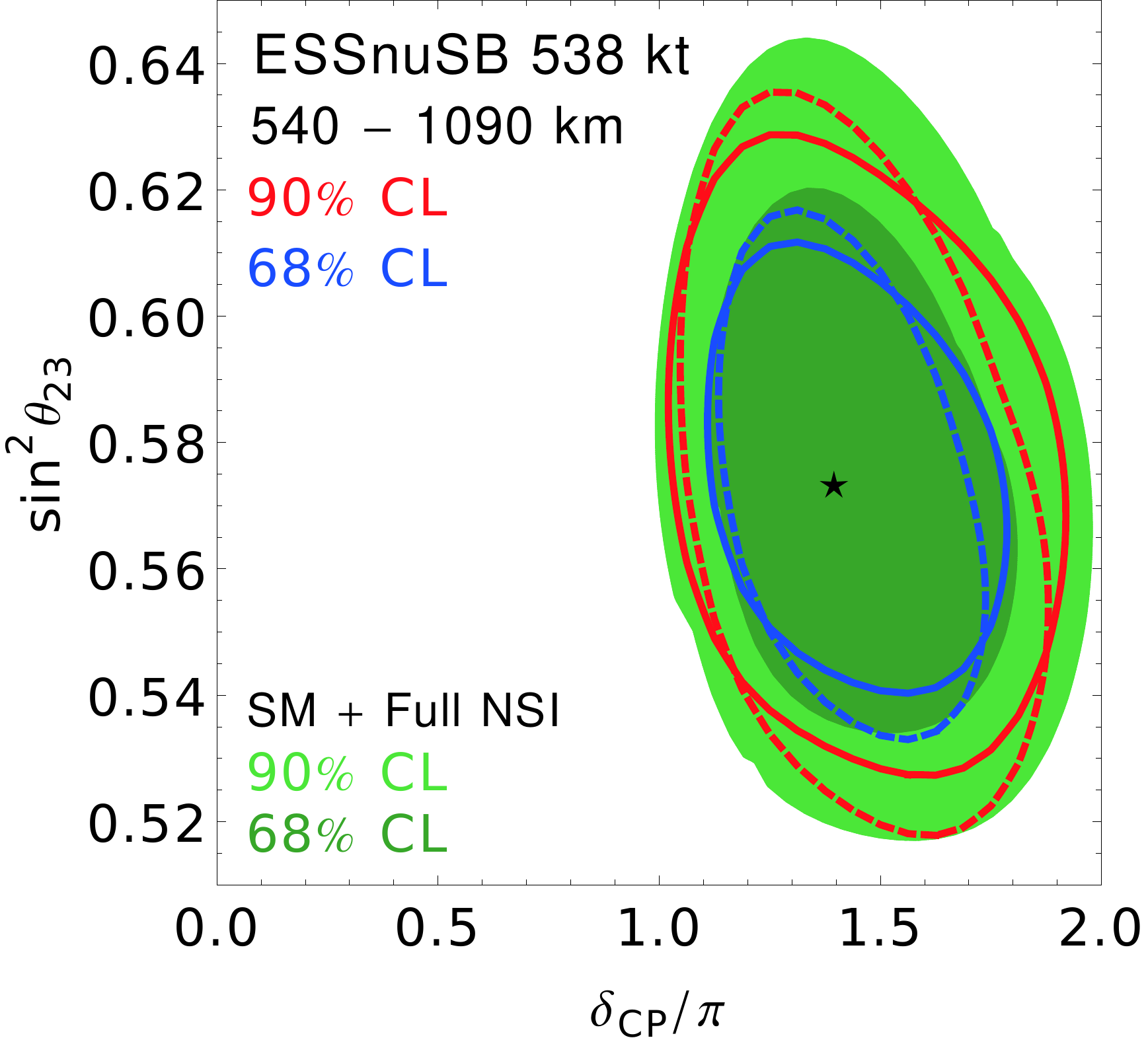}
               \end{subfigure}
		\hfill
  \begin{subfigure}[h]{0.31 \textwidth}
			\caption{  }
			\label{fb12b}
\includegraphics[width=\textwidth]
		{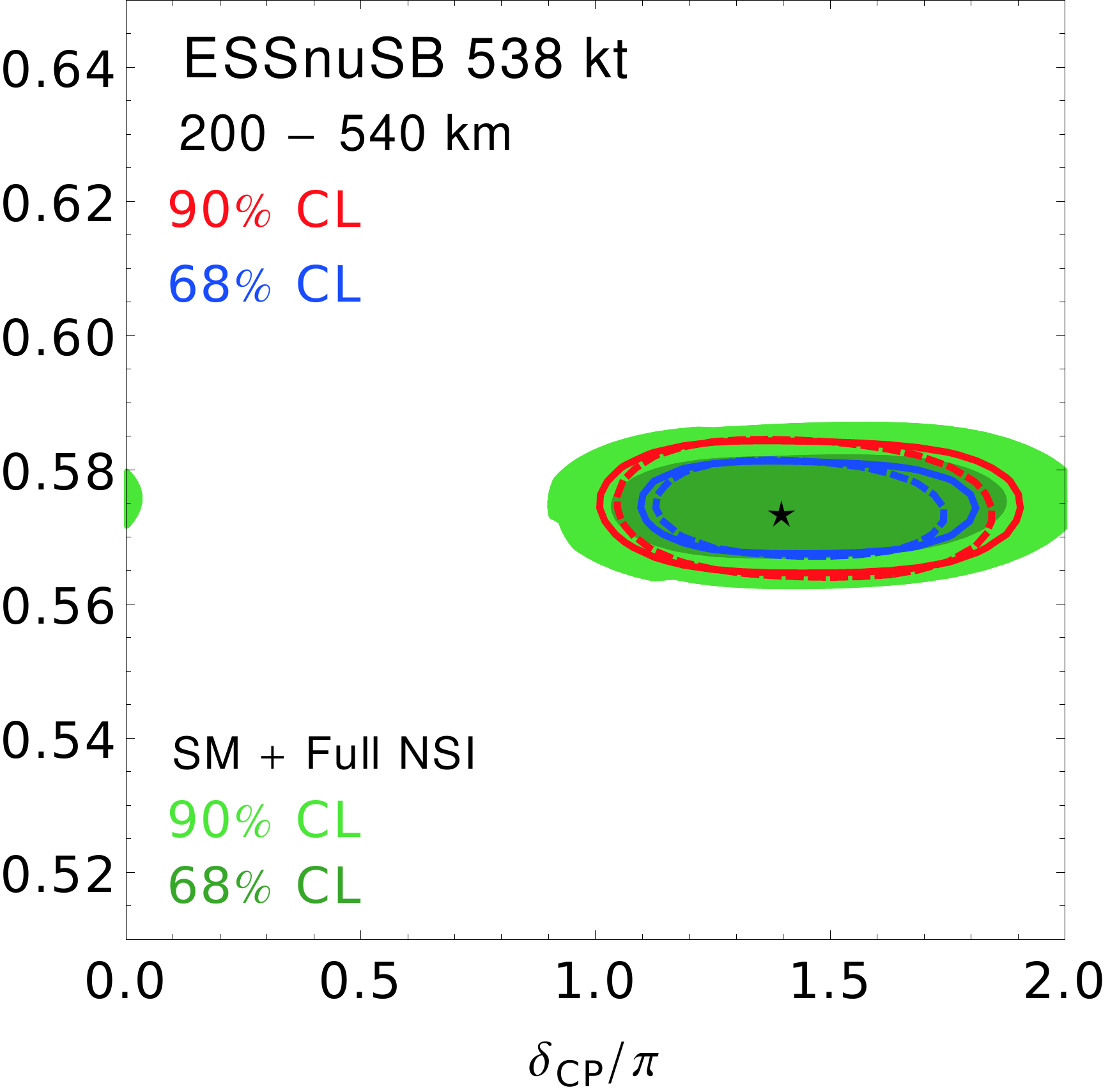}
               \end{subfigure}
		\hfill
		\begin{subfigure}[h]{0.31 \textwidth}
			\caption{}
			\label{fb23}
			\includegraphics[width=\textwidth]
		  {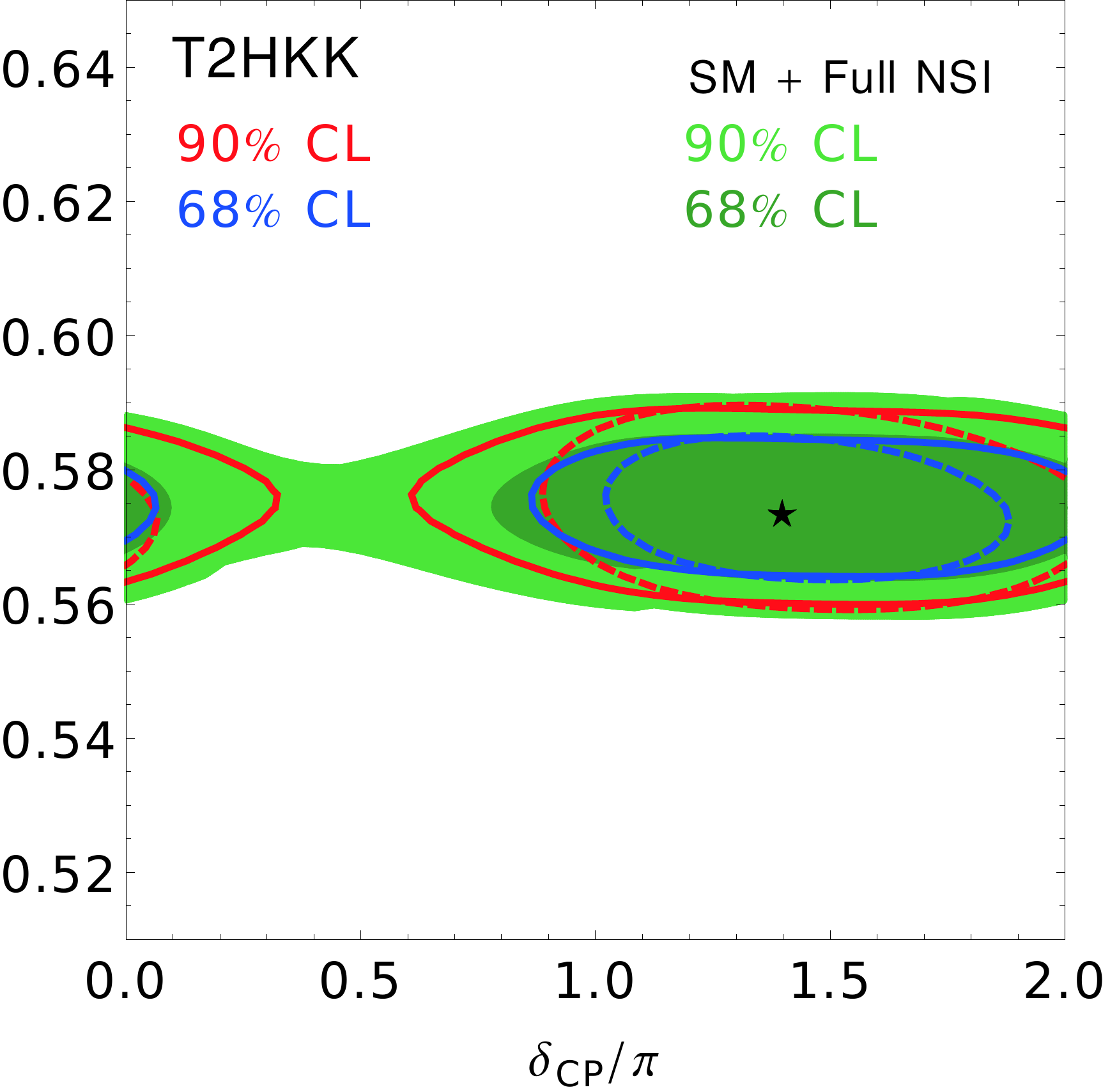}
            \end{subfigure}
		\hfill	
 \caption{Expected allowed regions in the ($\delta_{CP}$, $\sin^2 \theta_{23}$) plane. The standard 3$\nu$ oscillation framework (SM) is show in (solid lines) while (dashed lines) display the scenario with (SM + NSI) where we set ($|\epsilon_{e \mu}| =0.19$, $\phi_{e \mu}=1.5 \pi$), corresponding to the best fit values from Ref.~\cite{Denton:2020uda}. The left and middle panels display the ESSnuSB setup with two baselines at either 540$-$1090 km or 200-1090 km from the source, while the rightmost panel sets the T2HKK configuration. Finally, we display in green solid contours the expected allowed regions considering all the NSI entries (SM+Full NSI), see text for a detailed explanation. 
  }
  \label{fbs}
\end{figure}

In Fig.~\ref{fbs}, we introduce our results for the expected sensitivity at 68$\%$ C.L. and 90$\%$ C.L. in the ($\delta_{CP}$, $\sin^2 \theta_{23}$) plane. The left panel shows the ESSnuSB 540$-$1090 km setup, the middle panel displays the corresponding ESSnuSB 200$-$540 km configuration, and the rightmost panel shows the corresponding T2HKK setup. Besides the standard three neutrino mixing scenario, we have included
the effects of nonzero matter NSI parameters from the ($e-\mu$)
sector.

From the figure shown above, we can notice that, for the future
precision measurements of the Dirac $CP$-violating phase
$\delta_{CP}$, even in the presence of matter NSI, the
determination of $\delta_{CP}$ at ESSnuSB would not be
considerably affected. Therefore at ESSnuSB, the determined value of
$\delta_{CP}$ can be considered a faithful estimate of its true
value both in the SM and in the SM$+$NSI scenarios. Moreover, even if
the leptonic $CP$-phase $\delta_{CP} = (0~\text{or}~\pi)$,
ESSnuSB will be able to measure the possible NSI phases for large
enough values of the flavor changing parameter, $|\epsilon|$, as
shown, for example, in Fig.~3 of Ref.~\cite{Capozzi:2023ltl}. On the
other hand, for DUNE (right panel Fig.~\ref{fdune}) and T2HKK (rightmost
panel of Fig.~\ref{fbs}), the presence of matter
NSI modifies the determination of $\delta_{CP}$. As far as the
mixing angle $\theta_{23}$ precision is concerned, the inclusion of
matter NSI does not have a significant impact on its determination at
T2HKK. On the other hand, the determination of the mixing angle,
$\theta_{23}$, would be slightly affected at the ESSnuSB 540$-$1090 km
configuration.

\section{Conclusions}
\label{con}
This paper analyzed the determination of the leptonic $CP$-violating
phase, $\delta_{CP}$, in the presence of matter NSI from the
flavor-changing $e\mu$ and $e\tau$ sectors, as well as the incorporation of all the NSI parameters at several future LBL setups. We show that the ESSnuSB setup, located at 540 km or 360 km (a
practical vacuum oscillation experiment), will be able to determine a
faithful value of $\delta_{CP}$ regardless of matter NSI
effects. On the other hand, if the NOvA and T2K discrepancy on the
$CP$-phase measurement continues, a DUNE-like experiment, where matter
effects are significant, will be capable of determining the
corresponding matter NSI parameters with compelling precision, as
shown in Ref~\cite{Denton:2022pxt}. Moreover, we have illustrated that
DUNE will offer a superior sensitivity to the atmospheric mixing
angle, $\theta_{23}$, relative to the ESSnuSB configuration. We have
also shown that to obtain a reliable measurement of
$\delta_{CP}$, the combination of ESSnuSB and DUNE synergies
would be beneficial.

In addition, we investigated the constraining power for the leptonic $CP$-phase value at several experimental configurations with two baselines. The ESSnuSB 540$-$1090 km setup can contribute to the full physics program at ESSnuSB while being able to determine the Dirac $CP$-phase at good precision. Furthermore, the ESSnuSB 200$-$540 km configuration offers an opportunity to improve precision measurements on $\delta_{CP}$ as well as $\theta_{23}$, with respect to the single baseline ESSnuSB setup. Last but not least, within the aforementioned scenario, the T2HKK proposal will have a notable sensitivity to the matter NSI parameters. The restrictive power to the leptonic $CP$-phase may be modest in the configuration that we have studied. However, its determination of the atmospheric mixing angle $\theta_{23}$ is robust with respect to matter effects.

\section*{Acknowledgments}
We would like to thank L.~J.~Flores for his participation in early stages of this project.
This work was partially supported by SNII-M\'exico and CONAHCyT research
grant: A1-S-23238. We acknowledge the anonymous referee for the illuminating comments and suggestions.

%\bibliographystyle{unsrt}
%\bibliography{refs}

%\printbibliography

%%%%%%%%%%%%%%%%%%%%%%%%%%%%%%%%%%%%%%%%%%%%%%%%%%%%%%%%%%%%%%%%%%%%%%%
\end{document}